\documentclass[12pt]{iopart}

\usepackage{iopams}
\usepackage{graphicx}
\usepackage{color}
\usepackage{mathrsfs}

\begin{document}

\title[Self volume and decision-making in predator-prey system]{A single
predator charging a herd of prey: effects of self volume and predator-prey
decision-making}

\author{Maria Schwarzl$^\sharp$, Aljaz Godec$^{\sharp,\dagger}$, Gleb
Oshanin$^\ddagger$, and Ralf Metzler$^{\sharp}$}
\address{$\sharp$ Institute of Physics \& Astronomy, University of Potsdam, 14476
Potsdam-Golm, Germany\\
$\dagger$ National Institute of Chemistry, 1000 Ljubljana, Slovenia\\
$\ddagger$ Sorbonne Universit{\'e}s, UPMC Univ Paris 06, UMR 7600, LPTMC,
F-75005, Paris, France\\
E-mail: rmetzler@uni-potsdam.de}

\date{\today}

\begin{abstract}
We study the degree of success of a single predator hunting a herd of prey
on a two dimensional square lattice landscape. We explicitly consider the self
volume of the prey restraining their dynamics on the lattice. The movement of
both predator and prey is chosen to include an intelligent, decision
making step based on their respective sighting ranges, the radius in which
they can detect the other species (prey cannot recognise each other besides
the self volume interaction): after spotting each other the motion of prey
and predator turns from a nearest neighbour random walk into direct escape or
chase, respectively. We consider a large range of prey densities and sighting
ranges and compute the mean first passage time for a predator to catch a prey
as well as characterise the effective dynamics of the hunted prey. We find
that the prey's sighting range dominates their life expectancy and the predator
profits more from a bad eyesight of the prey than from his own good eye sight.
We characterise the dynamics in terms of the mean distance between the predator
and the nearest prey. It turns out that effectively the dynamics of this distance
coordinate can be captured in terms of a simple Ornstein-Uhlenbeck picture.
Reducing the many-body problem to a simple two-body problem by imagining predator
and nearest prey to be connected by a Hookean bond, all features of the model such
as prey density and sighting ranges merge into the effective binding constant.
\end{abstract}

\section{Introduction}

Every animal must eat in order to survive.  For certain predator species
this necessarily implies to chase and bring down a sufficient amount of prey.
With predators always on the lookout for food, prey must constantly be on
the alert.  While scattering and zigzagging to confuse the predator is a
popular method of herd animals to escape \cite{chen,olson}, if the escape
paths are not well co-ordinated individual prey may also block each other.
The self volume effect is also relevant in the hunt of killer cells
(macrophages, for instance) in biological organisms attacking bacteria
colonies or biofilms.\footnote{In the following we use the language of
predator-prey systems, keeping in mind the relevance of the model for such
cellular systems.}
In this paper we study the influence of self volume effects on a herd of
non-communicating prey with the autonomy of taking decisions on the run,
as quantified by the typical time to catch a prey.

In the study of the dynamics of predator-prey systems one is generically
interested in the likelihood for the survival of the prey as a function
of the parameters of the dynamics of both prey and predator. Prototype
mathematical models of predator-prey systems are reaction-diffusion models
\cite{scavenger,TargetAnnihilation,brey,Trapping,LatticeTrapping,SeaOfTraps},
in which both species are assumed to move randomly. In one dimension the
survival probability of a diffusing prey exposed to a number of diffusing predators
decays as a power law in time \cite{besieged,capture}. In two dimensions the
predators catch the prey with probability one, but the mean life time of the prey
is infinite.  The survival probability of a lamb in the presence of $N$ lions
in two dimensions decays logarithmically slowly as $\mathscr{S}_N(t)\sim\left(
\mathrm{ln}t\right)^{-N}$ \cite{capture}. In contrast, in dimensions $\ge3$ the
capture is unsuccessful
as a consequence of the transience of random walks \cite{Feller,Weiss}. Other
features considered in predator-prey models include finite life times of the
species \cite{Campos} or the presence of a third party in the form of a repellent
obstructing the predator to reach the prey \cite{Wang}. Moreover, three groups of
species hunting each other were modelled \cite{Sato}, owing to the fact that most
animal predators are prey of other animals themselves. Finally, effects of safe
havens for prey animals may be considered \cite{haven}.

While such continuum random walk models revealed various interesting results
it is clear that the escape and pursuit dynamics is at least partially
deterministic, that is, both predator and prey hunt or escape in some sense
intelligently.  A way to improve the mathematical modelling is to assume that
both species can see each other within a certain radius of vision and try to
use this as an advantage in the escape and pursuit process \cite{Oshanin,Sengupta}.
In such a model the motion consists of random walks which turn into directed
ballistic transport once predator and prey spot each other.  As shown in
\cite{Oshanin} the probability to escape can be greatly enhanced if the
prey can see the predator and has the possibility to run away.  During the
pursuit the prey's movement is superdiffusive. In this scenario a total of
three predators may be necessary to catch a single prey \cite{Oshanin}.
Predators may also optimise their search by sharing information
\cite{Martinez-Garcia}.
While the assumption of some level of intelligence certainly makes the model
more realistic, there is still one aspect that has up to now been ignored.
Namely, in reality prey are impenetrable bodies.  Thus, in an abundant
population of prey (a lion chasing a herd of antelopes, a wolf charging
at a flock of sheep, or a killer cell attacking a bacteria colony or biofilm)
the prey species may obstruct each other while trying
to escape.  The self volume (non-phantom) constraint greatly influences
the single species and collective dynamics of random walkers \cite{Bruna,Bruna2}
leading to qualitative differences in the walkers' motion.  Therefore, the
dynamics and survival probability in predator prey systems at intermediate
and higher prey densities is expected to be equally affected.  Recently a
herd of prey chased by a pack of predators including self volume effects
was studied \cite{aggregation}.  As a result the prey's survival time was
found to increase if the prey aim for a specific type of clustering.

In this paper we study the success of a single predator hunting a flock of
prey on a two-dimensional square lattice with periodic boundary conditions
taking into account the prey's self volume. In addition, both species
move intelligently in that they can influence their movement by visual
perception within their sighting range. The paper is structured as follows:
First we introduce our model. Next we present the numerical and analytical
results for the mean first capture time, which is the time the predator
needs to catch the first prey, as a function of prey density and the respective
sighting ranges. We find that
the mean first capture time as a function of prey density follows a power
law. The (non-universal) exponent depends on the sighting ranges of both
predator and prey. For the analytical calculations we split the predator's
motion into a diffusing part and a ballistic part, representing the search
for the prey and the direct chase, respectively. We then present a study
of the mean distance between predator and nearest prey, which is found to
decrease exponentially in time.  Using the mean distance we show that we can
capture its dynamics in terms of a simple Ornstein-Uhlenbeck process:
the relative motion of predator and nearest prey can thus effectively be
viewed to be a random process confined by an harmonic potential.
Neglecting all other prey, the model parameters such as
sighting ranges and prey density can be absorbed into the associated spring
constant.

\section{Lattice model}

\begin{figure}
\begin{center}
\includegraphics[width=8cm]{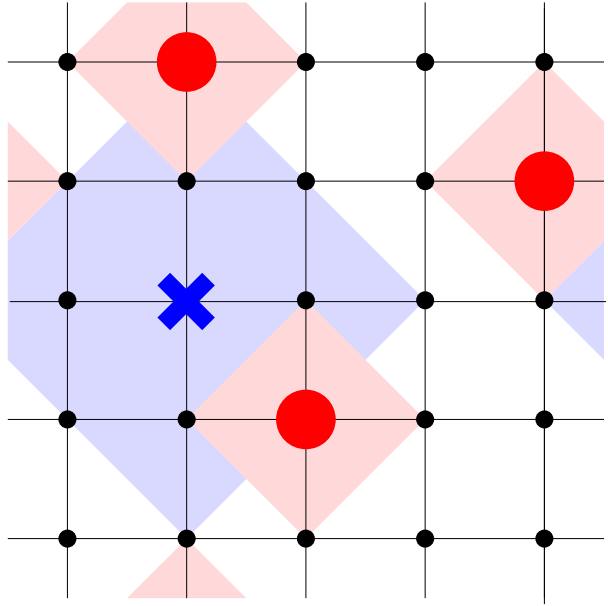}
\end{center}
\caption{Predator (blue cross) and prey (red dots) on a square lattice. The pale
blue and red diamonds represent their respective sighting ranges. Due to their
self volume different prey are not allowed to share the same lattice site. Once
a prey and the predator meet at the same lattice site the predator is considered
to have caught the prey.}
\label{model}
\end{figure}

To study the success of a single predator hunting a herd of prey we create an
agent-based simulation in which predator and prey move on a two dimensional
square lattice with periodic boundary conditions. Each species has its specific sighting range $\sigma$ in which it can see the other species as depicted in Fig \ref{model}.
Distances as well as sighting ranges are measured as chemical distances $d=\Delta
x+\Delta y$ of the added bond lengths,
with lattice spacing $a$ equal to unity.
The predator starts from the centre of the lattice and the prey are initially
randomly distributed---excluding the centre of the lattice---such
that the occupancy of a single site is less or equal to a single prey. Predators
and prey move
with the autonomy of decision in the following sense.  If no prey is in the
sighting range of the predator and, for a given prey, the predator is not in
its sighting range, both participants perform a nearest neighbour random walk.
If a prey comes into the sighting range of the predator, the predator chooses
a site randomly, subject to the condition that the distance $d$ to the prey
necessarily decreases.  Every lattice site that minimises the distance to
the prey is chosen with the same probability, lattice sites that increase
the distance cannot be chosen.  Analogously, if the predator is spotted the
prey chooses a site randomly, subject to the condition that the distance to
the predator necessarily increases.  If two or more prey are within the same
distance to the predator the latter chooses randomly which prey to pursue.
Due to the self volume of the prey, the prey's motion is restricted. 
In principle, there exist two possible ways to implement the self volume.
Either the prey chooses only from empty sites and always executes a jump as
long as there is at least one empty nearest neighbour site. Or the prey blindly
chooses a nearest neighbour site but only jumps if the chosen site is unoccupied;
otherwise, if it is occupied, the prey retains its location. We chose the latter
scenario, as this appears closer to the situation encountered for confused prey
or for moving bacteria. Using this update strategy, we simultaneously choose the
individual moves for the prey and the predator. In each round of motion updates
for the prey we randomly choose a sequence of individuals, thus avoiding any bias
among individuals \cite{Seitz}. According to this random sequence we then check
whether the individual prey are allowed to jump given the actual positions of
all other prey. The motion of the predator takes into account the positions of
all prey at the end of the previous update. Once all inidividual jumps of prey
and predator are determined, all positions of the entire predator-prey system
are updated simultaneously.

The time unit is chosen arbitrarily and relates to the diffusion constant $D$
\begin{equation}
\Delta t=\frac{a^2}{4D},
\end{equation}
where $a=1$ is the lattice spacing.\footnote{In these units, $D=1/4$ corresponds
to the diffusion coefficient for a single prey or predator moving on the lattice.}
After the individual steps of all participants are accomplished, we check if
the predator caught a prey.  If the first prey is caught the simulation terminates.
The mean first capture time and the mean distance are obtained from $10^4$
realisations and the first passage density is obtained from $10^6$ runs.

\section{Mean first capture time}
\label{section_MFCT}

We start by quantifying the success of the predator by computing the mean first
capture time $\left\langle \tau_c \right\rangle$, that is, the typical time the
predator needs to catch the first prey. In mathematical terms this corresponds
to the prey's survival time.  As one can easily imagine the mean first capture
time depends crucially on both sighting ranges $\sigma_{\mathrm{prey}}$
and $\sigma_{\mathrm{pred}}$ as well as on the prey density $\varrho=N/L^2$,
where $N$ is the number of prey and $L^2$ is the number of lattice sites.
A higher prey density reduces the prey's survival expectation.  One reason
is that the probability that initially one prey sits close to the predator
is higher and therefore the prey gets spotted earlier.  The second reason
is that a chased prey gets trapped more easily if there are more prey that
occupy nearest neighbour sites and therefore lead to a frustration of
the prey's mobility.

\begin{figure}
\begin{center}
\includegraphics[angle=-90,width=10cm]{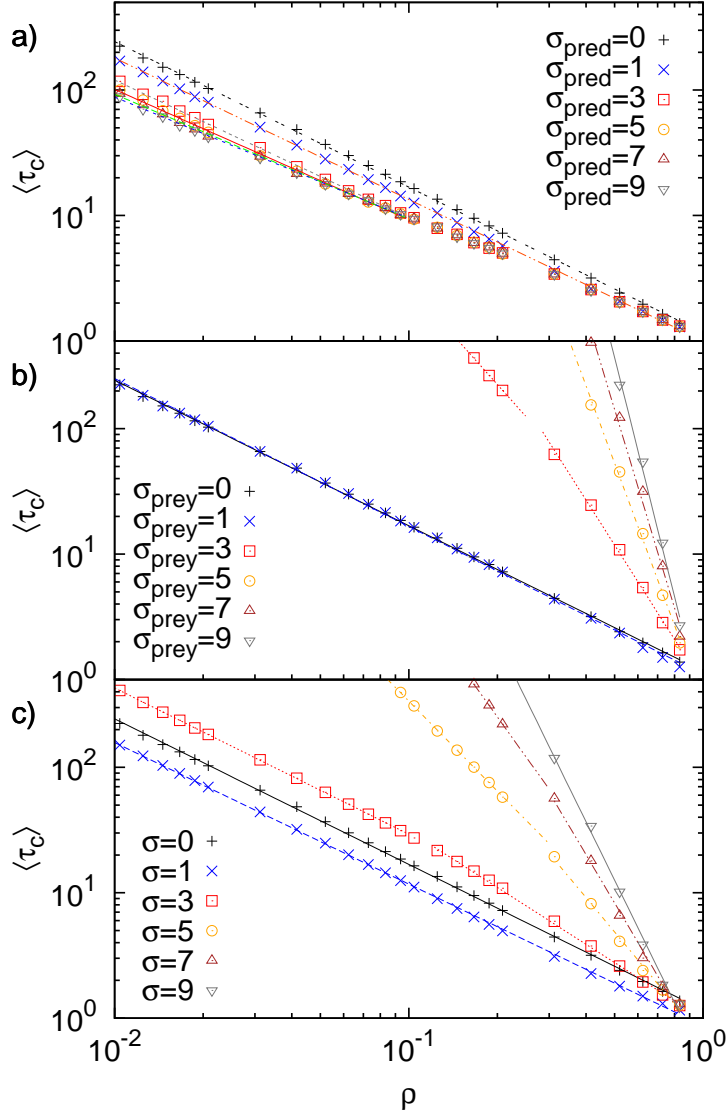}
\end{center}
\caption{Mean first capture time as a function of the prey density, averaged
over $10^4$ realisations. a) Blind prey, the predator's sighting range
increases from top to bottom: $\sigma_{\mathrm{pred}} =0,1,3,5,7,9$. b)
Blind predator, the preys' sighting range increases from bottom to top:
$\sigma_{\mathrm{prey}}=0,1,3,5,7,9$. c) Identical sighting ranges of prey and
predator, sighting ranges increase from bottom to top, $\sigma_{\mathrm{prey}}
=\sigma_{\mathrm{pred}}=0,1,3,5,7,9$.  The lines are power-law fits according
to Eq.~(\ref{Scaling_MFCT}).  The exponent $\beta$ as a function of sighting
ranges is depicted in Fig.~\ref{fig_beta_nu} in \ref{appb}.}
\label{fig:CaptureTime}
\end{figure}

In this setup we distinguish two limiting cases: A single prey ($\varrho=1/L^2$)
with sighting range greater than two can never get caught, its life time
is infinite.  Conversely, if every lattice spacing is occupied by a prey
($\varrho=1-1/L^2$) then the predator needs exactly one time step to catch the
first prey. For arbitrary densities, as a result of extensive simulations we find
from Fig.~\ref{fig:CaptureTime} that the mean first capture
time as a function of the prey density follows a power law behaviour
\begin{equation}
\left\langle\tau_c\right\rangle\sim\varrho^{-\beta\left(\sigma_{\mathrm{pred}},
\sigma_{\mathrm{prey}}\right)}
\label{Scaling_MFCT}
\end{equation}
in which different combinations of sighting ranges lead to different slopes.
Furthermore, there appears a crossover between two regimes for larger
sighting ranges of the prey, in which we find different slopes for the low
and intermediate density range and the high density range; see, for instance,
the square symbols in Figs.~\ref{fig:CaptureTime} b) and c).

\begin{figure}
\begin{center}
\includegraphics[width=8cm]{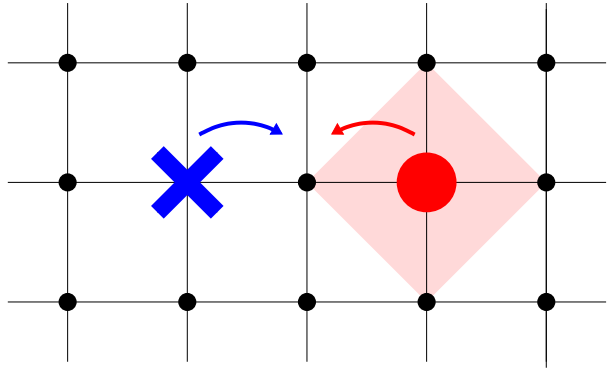}
\end{center}
\caption{A short-sighted prey ($\sigma_{\mathrm{prey}}$=1), depicted by the red
dot, can get caught by the predator (blue cross) despite his field of vision by
random collision, due to simultaneous jumps to the same lattice site.}
\label{fig:randomcollicion}
\end{figure}

In more detail, while the predator's sighting range only slightly influences the
prey's survival, as shown in Fig.~\ref{fig:CaptureTime}a), the prey can
increase their life expectancy significantly by a finite sighting range of at
least two,
compare Fig.~\ref{fig:CaptureTime}b), even in the case of a long sighting
range of the predator, see Fig.~\ref{fig:CaptureTime}c).  In both figures
\ref{fig:CaptureTime}b) and \ref{fig:CaptureTime}c) a significant variation
at intermediate  $\sigma$ values is distinct.  We note that a short sighting
range of the prey ($\sigma_{\mathrm{prey}}=1$) has no advantage over a
vanishing one. An explanation can be found ``microscopically''.  There are
two possibilities for a prey to get caught. First, a prey gets stuck and,
despite his eyesight, cannot evade the encounter with the predator; or,
second, predator and prey simultaneously jump on the same lattice site and
collide randomly, see Fig.~\ref{fig:randomcollicion}.  With sighting range
zero or one a prey cannot foresee a random collision, because the distance
decreases instantly from two to zero.  Thus, the prey needs at least a
sighting range of two to prevent such a situation.  These random collisions
further lead to the fact that the predator is more successful with an even
sighting range $\sigma_{\mathrm{pred}} = 2n$, $n$ being an integer number,
than with a higher odd one $\sigma_{\mathrm{pred}} = 2n + 1$.  Since the
random collision is a natural and frequent way to get caught, we decided to
eliminate these effects by treating only odd sighting ranges.

We note that we did not include error bars in our figures.
A stochastic variable with exponential (Poissonian) probability density function
$ p(t)=\tau^{-1}\mathrm{e}^{-t/\tau}$ has the mean $\tau$ and variance $\tau^2$. 
The mean first capture time presented in this section is the first moment of the 
exponentially distributed first passage density obtained in section \ref{section_density}.
The standard deviation of this Poissonian process $\sigma=\sqrt{\int_0^\infty t^2p(t)-\tau^2\mathrm{d}t}$ is equal to the mean $\tau$, which is indeed confirmed from
our numerical results with a sample size of $10^4$ per data point. Repeated
simulations produced practically indistinguishable results.

\section{Distribution of first capture times}

In comparison to an ensemble of non-interacting random walkers self volume
effects and the autonomy of decision-making of the participants limit the
possibilities of analytical calculations.  We succeeded in calculating the
distribution of the time for catching the first prey only in the case of
blind prey.  As the results are nevertheless instructive we discuss this
case here in some detail. The autonomy to switch the mode of motion of the
predator can be included by dividing the process into two subprocesses.
The first one describes the diffusing predator while looking out for a prey.
The second subprocess portrays the direct chase of the prey, which can in
fact be considered as a ballistic motion in chemical space such that the
predator still has the option of choosing sites in different directions.

\subsection{Searching the prey}
\label{Searching}

The first subprocess describes the random motion of the predator while looking
out for a prey.  According to the model during that time the predator performs
a nearest neighbour random walk on the lattice.  We are interested in the
first passage density function of the predator to find the first prey, that is,
until the first prey enters the predator's sighting range.  For simplification
we use a continuous radial coordinate and ignore the fact that the participants
move on a lattice. We assume that there exists an effective radius $r_{\mathrm{
eff}}$ around the predator in which he will not encounter a prey.
This radius has a natural lower bound which is the initial distance between
predator and nearest prey (at time $t=0$), calculated in section
\ref{meandistanceanalytics}.  If the predator hits this effective radius,
he spots the prey and will from there on switch his motion to the direct
chase calculated in the next subsection.

We consider the predator as a diffusing particle in two dimensions
and calculate his first passage time to escape a sphere with radius
$r_{\mathrm{\mathrm{eff}}}-\sigma_{\mathrm{pred}}$.  For simplification we let the
particle diffuse between concentric spheres with an inner reflecting boundary
at radius $R_-$, which will later tend to zero, and an outer absorbing boundary
at radius $R_+$, representing the point where the predator spots a prey.  $R_+$
is thus the distance between predator and prey minus the sighting range of the
predator. The predator starts inside the interval $R_-<r_0<R_+$. We will later
let $r_0$ tend to $R_-$ to capture the predator's starting position correctly.
The diffusing particle can be described by the radial diffusion equation
\begin{equation}
\frac{\partial p(r,t)}{\partial t}=D\frac{1}{r^2}\frac{\partial}{\partial r}
\left(r^2\frac{\partial}{\partial r}\right)p(r,t)
\end{equation}
for the probability density function $p(r,t)$ to find the predator at radius $r$
at time $t$. The initial condition we choose as $p(r,t=0)=\delta(r-r_0)/\left(2\pi
r_0\right)$, that is, the particle starts at $r=r_0$. We impose the absorbing
boundary condition $p(R_+,t)=0$ at $R_+$ and the reflecting boundary condition
$-[\partial p(r,t)/\partial r]_{R_-}=0$ at $r=R_-$. After Laplace transform 
\begin{equation}
\tilde{f}(s)=\mathscr{L}\{f(t)\}(s)=\int\limits_0^{\infty}f(t)e^{-st}\mathrm{d}t
\end{equation}
and with $x=r \sqrt{s/D}$ the diffusion equation is reduced to the ordinary
differential equation
\begin{equation}
\tilde{p}(x,s)-\frac{1}{x} \frac{\partial \tilde{p}(x,s)}{\partial x}-\frac{
\partial^2 \tilde{p}(x,s)}{\partial x^2}=\frac{1}{D}\frac{\delta(x-x_0)}{2\pi x_0}
\end{equation}
For $x<x_0$ and $x>x_0$ this is the modified Bessel equation of zero order
with known solution $\tilde{p}(x,s)=C_1 I_0(x) + C_2 K_0(x)$ for $x\neq x_0$
\cite{Watson}, where $I_0(x)$ and $K_0(x)$ are the modified Bessel functions
of first and second kind.  We solve this equation by imposing the continuity
condition $\tilde{p}_<(x_0,s)=\tilde{p}_>(x_0,s)$ and the jump-discontinuity
\begin{equation}
\left.-\frac{\partial\tilde{p}_>(x,s)}{\partial x}\right|_{x_0}+\left.\frac{
\partial\tilde{p}_<(x,s)}{\partial x}\right|_{x_0}=\frac{1}{2\pi Dx_0}
\end{equation}
where $\tilde{p}_<(x,s)$ is the solution in the range $x<x_0$ and $\tilde{p}_>
(x,s)$ is the solution in the range $x>x_0$. With the shorthand notations $C_\nu
(a,b)=I_\nu(a)K_\nu(b)-K_\nu(a)I_\nu(b)$ and $D_{\nu,\pm}(a,b)=I_\nu(a)K_{\nu\pm1}
(b)+K_\nu(a)I_{\nu\pm1}(b)$ \cite{redner} the solution yields in the form
\begin{equation}
\tilde{p}(x,s)=\frac{C_0(x,x_+)}{2\pi D x_0\left(\frac{C_0(x_0,x_+)C_{-1}(x_0,x_-)}
{D_{0,-}(x_0,x_-)}-D_{-1,+}(x_0,x_+)\right)}.
\end{equation}
If the particle starts at the inner boundary $r_0=R_-$, corresponding to $x_-$
in the reduced coordinates,
\begin{equation}
\lim_{x_0\rightarrow x_-}\tilde{p}_>(x,s)=-\frac{C_0(x,x_+)}{2\pi Dx_0 D_{-1,+}(
x_0,x_+)}
\end{equation}
we calculate the flux through the outer boundary as
\begin{equation}
-2\pi x_+D\left.\frac{\partial\tilde{p}_>(x,s)}{\partial x}\right|_{x_+}=
\left(x_- D_{-1,+}(x_-,x_+)\right)^{-1}.
\end{equation}
When $x_-$ approaches zero, we therefore find that
\begin{equation}
\lim_{x_-\rightarrow0}\tilde{\wp}_{\mathrm{search}}(s)=\left(I_0(x_+)\right)^{
-1}=\left(I_0\left(R_+\sqrt{s/D}\right)\right)^{-1},
\label{sub1}
\end{equation}
where on the right hand side we restored the original variables.  This is
but the first passage time density function in Laplace space of the predator
to spot a prey.  From that time the predator will chase the prey directly,
this part being calculated in the next subsection.

\subsection{Chasing the prey}
\label{DirectChase}

The second subprocess, which describes the predator's movement from the
moment of spotting the prey until the prey is caught, can be reduced to a
one-dimensional problem.  Remember that the decision for every step of the
predator is constrained by the following rule: the distance to the prey
has to necessarily decrease.  For the prey, analogously, the goal is to
increase the distance.  Consequently after a combined predator and prey
step the distance between predator and prey
can either stay the same or decrease by one lattice spacing if the chosen
site of the prey is already occupied and the prey remains at its site.
The first capture time can thus be calculated exactly from the number of
times a prey remains at its location.  A large sighting range of the prey
renders the analysis of the chasing process more difficult as all prey try
to escape from the predator
and will eventually build a cluster that moves away from the predator.
Due to the random order of the updates, one cannot say which of the prey
remains sitting.  Therefore, we confine ourselves to the case of blind prey.
In this case the prey undergoes normal diffusion and the predator moves
constantly towards the prey.  We therefore consider the predator to be a
moving cliff towards a diffusing particle, the blind prey.  The survival
probability of a diffusing particle in presence of a ballistically moving
cliff decays exponentially \cite{redner},
\begin{equation}
\mathscr{S}(t)\simeq (t/\tau)^{-1/2}e^{-t/\tau}.
\label{cliff}
\end{equation}
The associated first passage density $\wp(t)=-\mathrm{d}\mathscr{S}(t)/
\mathrm{d}t$ then becomes
\begin{equation}
\wp_{\mathrm{chase}}(t)\simeq e^{-t/\tau}\left(\frac{(t/\tau)^{-1/2}}{\tau}+\frac{(t/\tau)^{-3/2}}
{2\tau}\right).
\end{equation}
With the Laplace transform $\tilde{\wp}_{\mathrm{chase}}(s)\sim\left(1+\tau s\right)^{-1/2}$
in the long time limit corresponding to a small $s$ expansion, we finally get
$\tilde{\wp}_{\mathrm{chase}}(s)\sim\left(1+\tau s/2\right)^{-1}$.

Using the first passage time densities of the subprocesses of search and chase
we calculate the total first capture time density function in the next subsection.

\subsection{Density of first capture time}
\label{section_density}

\begin{figure}
\begin{center}
\includegraphics[angle=-90, width=10cm]{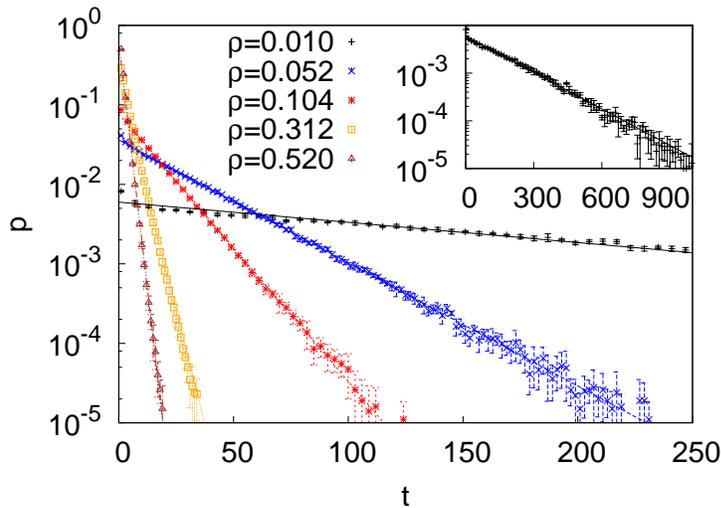}
\end{center}
\caption{First passage density for the case of blind prey and a short-sighted
predator ($\sigma_{\mathrm{pred}}=1$) for different prey densities. Each data
point shows the mean result from $10^6$ realisations. The error bars were
computed from splitting up the $10^6$ independent runs into ten runs of $10^5$
runs. Inset: Same plot for prey density $\varrho=0.052$ on a
larger scale. The lines are exponential fits according to
Eq.~(\ref{eq_scaling_density}).  The exponent $\lambda$ as a function of
prey density is depicted in Fig.~\ref{fig_lambda} in \ref{appb}.}
 \label{FP}
\end{figure}

The distribution of the first capture time is now given by the convolution
of results $\tilde{\wp}_{\mathrm{search}}(s)=(I_0(R_+\sqrt{s/D}))^{-1}$
and $\tilde{\wp}_{\mathrm{chase}}(s)\sim\left(1+\tau s/2\right)^{-1}$,
\begin{equation}
\wp(t)=\int_0^t\wp_{\mathrm{search}}(t')\wp_{\mathrm{chase}}(t-t')\mathrm{d}t' 
\end{equation}
which designates the probability that the predator spots the first prey at
time $t'$ and catches the prey during the time span $t-t'$.  In Laplace space
this convolution simplifies to the product $\tilde{\wp}(s)=\tilde{\wp}_{\mathrm{
search}}(s)\tilde{\wp}_{\mathrm{chase}}(s)$. The inverse Laplace transform can
be obtained in the long
time limit, corresponding to taking $s\rightarrow 0$. We thus need to invert
\begin{equation}
\tilde{\wp}(s)\simeq\left(1+(\kappa+\lambda)s+\kappa\lambda s^2\right)^{-1},
\end{equation}
where $\kappa=R_+^2/(4D)$ and $\lambda=\tau/2$. Taking the leading terms for
small $s$, $\wp(t)\simeq\mathscr{L}^{-1}\left\lbrace\left(1+\Lambda s\right)
^{-1}\right\rbrace$ with $\Lambda=\kappa+\lambda$, the inverse Laplace
transform yields the final result,
\begin{equation}
\wp(t)\simeq\Lambda^{-1}e^{-t/\Lambda}.
 \label{eq_scaling_density}
\end{equation}
This density of first capture is thus an exponential distribution, where
the rate $\Lambda^{-1}$ is a function of the prey density and the predator's
sighting range.  Fig.~\ref{FP} shows the numerical data of our simulation
for the case of blind prey and a short sighting range of the predator ($
\sigma_{\mathrm{pred}}=1$). The exponential form (\ref{eq_scaling_density}) agrees
quite well with the data over the whole density range.

\begin{figure*}
\begin{center}
\includegraphics[angle=-90,width=16cm]{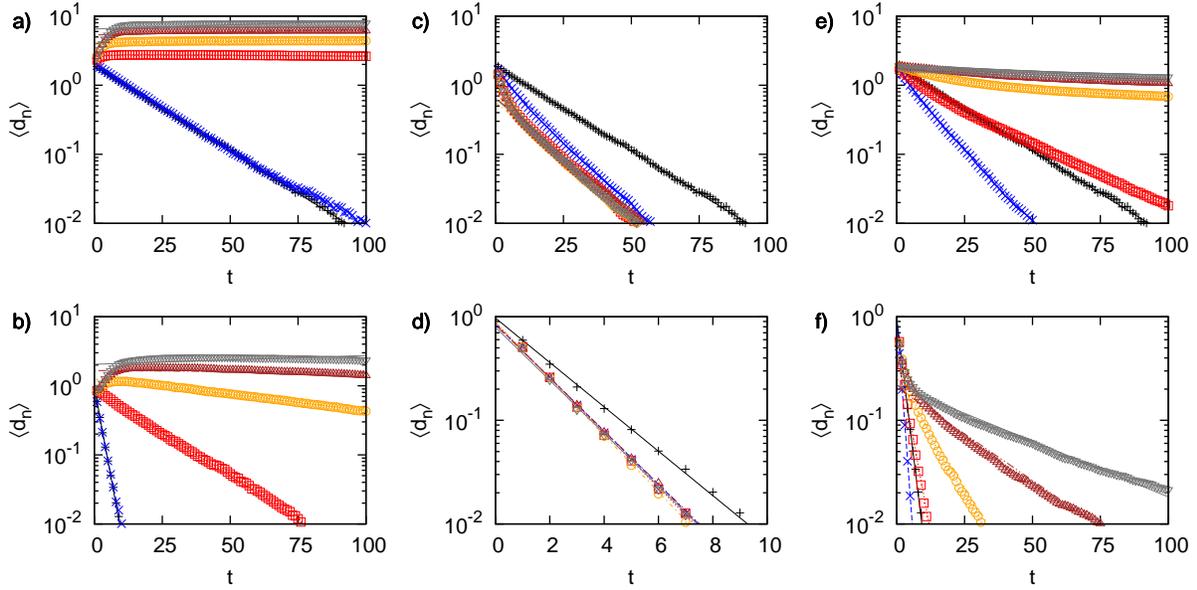}
\end{center}
\caption{Mean distance between the predator and the nearest prey as function
of time, averaged over $10^4$ realisations. The upper row (panels a, d, e) shows
the case of a low density
$\varrho=0.104$ and the lower row (panels b, d, f) represents the case of an
intermediate
density $\varrho=0.520$.  The two left panels a) and b) represent the case of
a blind predator, the preys' sighting range decreases from top to bottom
($\sigma_{\mathrm{prey}}=9,7,5,3,1,0$). The two middle panels c) \& d)
represent the case of blind prey. The predator's sighting range increases
from top to bottom ($\sigma_{\mathrm{pred}}=0,1,3,5,7,9$).  The two panels
on the right e) \& f) show the mean distance in case of identical sighting
ranges ($\sigma_{\mathrm{pred}}=\sigma_{\mathrm{prey}}=0,1,3,5,7,9$).
They decrease from top to bottom. 
The lines are exponential fits according to Eq.~(\ref{eq_scaling_meandistance}).}
\label{Meandistance}
\end{figure*}

\section{Mean distance between predator and nearest prey}

We now turn to study the dynamics of the mean distance between the
predator and the nearest prey in more detail.  In Fig.~\ref{Meandistance}
our simulation results for this mean distance are plotted for a low prey
density $\varrho=0.104$ in the upper row (panels a), c) and e)) and for an
intermediate density $\varrho=0.520$ in the lower row (panels b), d) and
f)). The distance decreases exponentially in time except for the case when a
blind predator is combined with a low prey density (Fig.~\ref{Meandistance}
a)) or with a very good eye-sight of the prey (Fig.~\ref{Meandistance} b)). In
these cases the distance is approximately constant in the shown time window.
In case of identical sighting ranges the distance between short-sighted species
decreases faster than the distance between blind species.  This phenomenon
is due to the random collisions explained in section \ref{section_MFCT}.

As intuitively expected, the distance between the predator and the nearest prey
decreases faster in the case of a large sighting range of the predator. However
when the prey's sighting range is large it softens the decay of the distance.
A high prey density also leads to a faster decay of the distance between the
predator and the nearest prey, because it implies more prey-prey obstruction
events for the chased prey, and with every one such event the distance is reduced
by one lattice spacing.

A naive model that captures the effective interaction between two diffusive
particles such as the predator and the nearest prey turns out to be the
Ornstein-Uhlenbeck process \cite{OrnsteinUhlenbeck}. It is defined in terms of
the stochastic differential equation
\begin{equation}
\mathrm{d}x(t)=(\textcolor{blue}{e}-cx(t))\,\mathrm{d}t+b\,\mathrm{d}W(t)
\end{equation}
with non-negative parameters $e$, $b$, and $c$. $W(t)$ denotes the Wiener process
\cite{vankampen}. The Ornstein-Uhlenbeck process describes the relaxation of the
variable $x$ with initial value $x(t=0)=x_0$ to the mean value $e/c$ in the
presence of Gaussian white noise. The first moment is given by the exponential
decay
\begin{equation}
\left\langle x(t)\right\rangle=\frac{\textcolor{blue}{e}}{c}\left[1-\exp(-ct)\right]+x_0\exp(-ct).
\end{equation}
Comparing the first moment to the observed simulated decay of the mean distance
between the predator and the nearest prey (Fig.~\ref{Meandistance}),
\begin{equation}
 \left\langle d_n\right\rangle\simeq
\mathrm{e}^{-\theta t},
\label{eq_scaling_meandistance}
\end{equation}
we see that the mean distance decreases as a special case of the Ornstein-Uhlenbeck
process with vanishing excentricity parameter, $e=0$.

\begin{figure}
\begin{center}
\includegraphics[angle=-90,width=10cm]{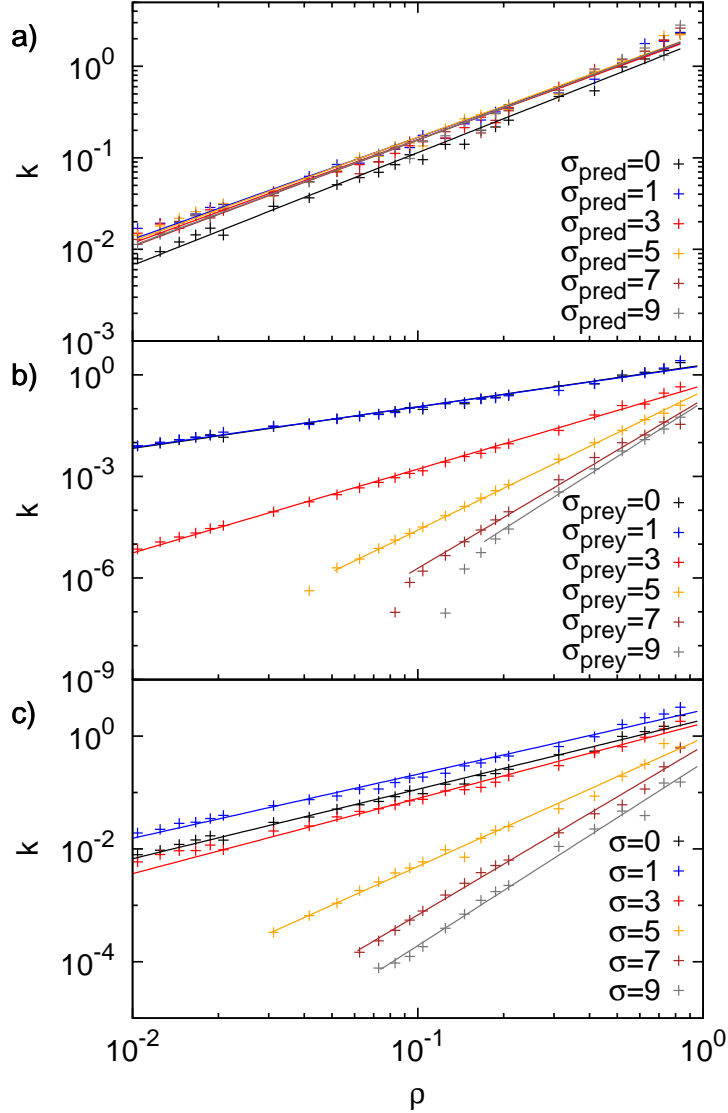}
\end{center}
\caption{Effective spring constant $k$ of our Ornstein-Uhlenbeck model as function
of the prey density for a) blind prey: $\sigma_{\mathrm{pred}}=0,1,3,5,7,9$,
increases from bottom to top. b) blind predator: $\sigma_{\mathrm{prey}}=0,1,3,5,7,
9$, increases from top to bottom. c) identical sighting ranges: $\sigma_{\mathrm{
prey}}=\sigma_{\mathrm{pred}}=0,1,3,5,7,9$, increases from top to bottom. 
The lines are power law fits according to Eq.~(\ref{eq_scaling_springconstant}). 
The exponent $\nu$ as a function of sighting ranges is depicted in
Fig.~\ref{fig_beta_nu} in \ref{appb}.}
\label{SpringConstants}
\end{figure}

A popular application of the Ornstein-Uhlenbeck process in physics is a Hookean
spring with spring constant $k$, whose dynamics is highly overdamped with friction
coefficient $\gamma$ in the presence of thermal fluctuations.  Therefore we
can imagine the predator and the nearest prey to be connected by a Hookean spring
and being driven by an external Wiener noise. The corresponding mean relaxes to
zero. The equilibrium length of the spring is therefore zero. The bottom of the
corresponding harmonic potential thus represents the capture of the prey by the
predator. Due to the analogy, the respective sighting ranges and the prey
density affect the stiffness of the spring. The spring constant is easily related
to the decay rate $\theta$ of the mean distance, $\theta(\varrho,\sigma)=k(\varrho,
\sigma)/2$. As shown in Fig.~\ref{SpringConstants} the fitted values for the
spring constant display the power law behaviour 
\begin{equation}
k\sim\varrho^\nu.
\label{eq_scaling_springconstant}
\end{equation}
The spring constant corresponds to the slopes of the functions in
Fig.~\ref{Meandistance}, extracted from the exponential fit and plotted as a
function of the prey density. It relates to the mean first capture time discussed
in section \ref{section_MFCT} in the following way. The exponential decay of the
mean distance between predator and nearest prey has a mean life time related to the
decay rate
\begin{equation}
\tau=\frac{1}{\theta}.
\end{equation}
Since the nearest prey is the one that will get caught, its mean first capture
time is related to the mean life time of the mean distance and consequently to
the inverse of the decay rate, compare Figs.~\ref{fig:CaptureTime} and
\ref{SpringConstants}.

\subsection*{Initial distance analysis}
\label{meandistanceanalytics}
 
We finally mention an analytical approximation for the distance between the predator
and the nearest prey. Since we want to capture the whole dynamics we first
need to determine the initial distance between predator and nearest prey at time
$t=0$. In the simulation we place the predator in the centre and place the prey
randomly around him including the self volume interaction. Then we measure the
distance between the predator and the nearest prey.
In section \ref{Searching} we used an effective radius $r_{\mathrm{\mathrm{eff}}}$
within which the predator does not encounter a prey.  Although we cannot calculate
this effective radius a natural lower bound is the initial mean distance
$\left\langle d_n\right\rangle_{t=0}$ between the predator and the nearest
prey.  Within this distance there is no prey present and therefore it is
impossible for the predator to encounter a prey.

We determine the initial distance between the predator and the nearest prey
on a square lattice with edge length $L$.  The predator sits in the
centre of the lattice and the prey are randomly distributed on the remaining
$N_S=L^2-1$ sites.  As the prey have a self volume, a lattice site can only
be occupied by a single prey.  The probability for the distance between predator
and nearest prey $d_n$ to be equal to $d$ is
\begin{equation}
P(d_n=d)=P(d_n\geq d)-P(d_n\geq d+1).
\label{eq_Appendix1}
\label{eq1}
\end{equation}
We then calculate the probability $P(d_n\geq d)$ using combinatorics. The
detailed calculation can be found in Appendix A. For the probability
function of the distance between predator and nearest prey we obtain
\begin{equation}
P(d_n)=\left[{N_R(d) \choose N_P}-{N_R(d+1) \choose N_P}\right]\Big/\sum\limits_{
i=1}^{d_{\mathrm{max}}}{N_R(d_i) \choose N_P},
\label{eq_pdf_initial}
\end{equation}
where we define $d_{\mathrm{max}}$ as the maximal possible distance between the predator
and the nearest prey. The expectation value of the initial distance from the
predator to the nearest prey, $\left\langle d_n\right\rangle=\sum\limits_{d_i=1}^{
d_{\mathrm{max}}} p(d_{\mathrm{min},i})d_{\mathrm{min},i}$ then yields in the form
\begin{equation}
\left\langle d_n\right\rangle=\frac{\sum\limits_{d_i=1}^{d_{\mathrm{max}}}d_i{N_R(
d_i) \choose N_P}-{N_R(d_i+1) \choose N_P}}{\sum\limits_{d_i=1}^{d_{\mathrm{max}}}
{N_R(d_i) \choose N_P}-{N_R(d_i+1) \choose N_P}}.
\label{eqmeandistance}
\end{equation}

\begin{figure}
\begin{center}
\includegraphics[angle=-90,width=10cm]{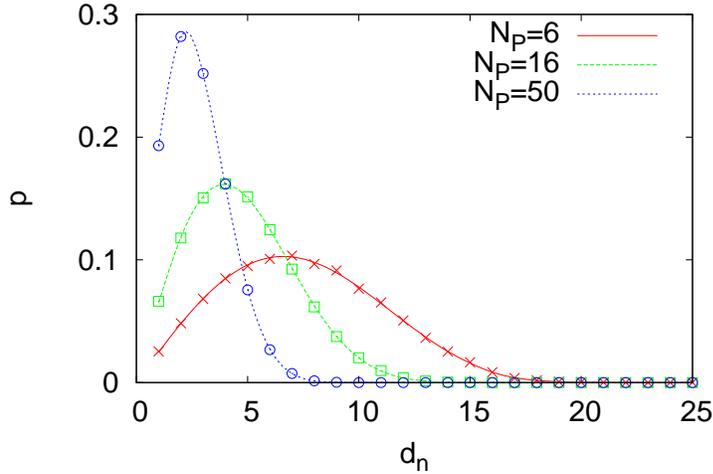}
\end{center}
\caption{Probability distribution of the initial distance between the
predator and the nearest prey for the case of different numbers $N_P$ of prey on
a square lattice with edge length $L=31$. The crosses represent the numerical data,
averaged over $10^4$ realisations. The lines show the analytical result
(\ref{eq_pdf_initial}).}
\label{fig_dmin}
\end{figure}

\begin{figure}
\begin{center}
\includegraphics[angle=-90,width=10cm]{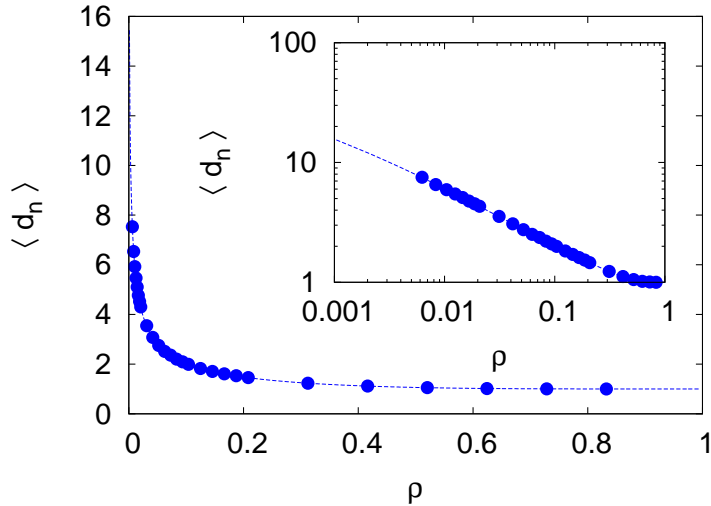}
\end{center}
\caption{Initial value of the mean distance between the predator
and the nearest prey as a function of prey density on a square lattice with edge
length $L=31$. The symbols represent the numerical data, averaged over $10^4$
realisations, and the dashed line the analytical result (\ref{eqmeandistance}).}
\label{fig_meandistance}
\end{figure}

The probability distribution of the initial distance to the nearest
prey is shown in Fig.~\ref{fig_dmin} and the related initial mean distance
as a function of prey density can be seen in Fig.~\ref{fig_meandistance}.
We simulated both the initial distance distribution and the initial
mean distance between the predator and the nearest prey by placing all
participants on the lattice under the model conditions with $10^4$ iterations.
Both analytical and numerical results show excellent agreement
in Figs.~\ref{fig_dmin} and \ref{fig_meandistance}.

\section{Discussion}

We studied the predator-prey dynamics of a single predator hunting a herd of prey
on a square lattice with decision-making species. While many predator-prey models
deal with collective predation \cite{Kamimura,Iwama,Angelani,Nishi,Ramanantoanina}
or the search for the optimal number of predators given the number of prey
\cite{Vicsek}, we chose a model consisting of one predator and many prey, which
is often found in Nature. Solitary hunters such as tigers,
bears, or sea turtles often have herd animals as their target. A tiger, for example,
hunts a herd of antelopes or a flock of sheep, a bear fishing a salmon out of a
swarm, or sea turtles eating jellyfish, shrimp, and fish living in schools.
Similarly individual killer cells in biological organisms may attack a colony of
bacteria or a biofilm.

A major ingredient of our model is the self volume of the prey, such that
no two prey are allowed on a given lattice site. We showed that in the case
of impenetrable prey the predator hunts more successfully if the prey have
worse eyesight. Moreover, we found that the predator benefits more from
a deterioration of the prey's eyesight than from an improvement of his
own eyesight.

While trapping reaction models obtain a minor influence of the prey's long
time survival probability by their diffusion constant \cite{brey,Trapping}
we found the prey's sighting range and thereby motion predominating their
survival probability.  Due to self volume interactions the prey are forced
to improve their eye-sight, and with a good field of vision can drastically
increase their chances of survival even in the range of high densities.

The prey only profit from a sighting range of at least two.  A very short
eyesight does not at all improve its survival probability with respect to being
blind.  This is attributed to random collisions between predator and prey.
Using a simplified analytic approach we showed that in the long time limit
the first passage density of the predator to catch a blind prey decays
exponentially in time with a non-linear dependence of the decay rate on the
prey density.

The effective motion during the chase (described in terms of the distance
between the predator and the chased prey) can be effectively described as a
linear relaxation process in an harmonic potential with a stochastic driving
where the density and sighting ranges determine the stiffness of the
corresponding Hookean spring. All non-linear effects entering
the motion due to self volume interactions can thus effectively be described
with a single parameter.

There exist a range of further open questions. To imitate natural environment one
could extend the dynamics by introducing (time or sighting range dependent)
waiting times.  One could choose different rates of motion for predator and
prey as well or even distribute the rates within the prey to simulate old,
sick or infant animals.  Additionally, many prey live in herds, so one could
let the prey be clustered as the initial condition.  Last but not least,
communication between the prey is a reasonable thing to assume.  Once one of
the prey spots the predator, immediately all of them are informed (similar
to stamping of rabbits or the cheeping calls of groundhogs), that is, a
collective response of prey.

We finally note that random search processes with non-Brownian search dynamics
are also widely discussed in literature.
While Brownian motion is an advantageous process to
find nearby targets \cite{Palyulin}, it is known that pure stochastic motion
leads to oversampling of the area on longer time scales. Hence, the optimal
number of encounters with prey can be found by switching between search
modes \cite{bartumeus,bartumeus2}.  Representative for such a process is
for example the intermittent search strategy which combines phases of slow
motion, allowing the searcher to detect the target, and phases of fast motion
during which targets cannot be detected \cite{intermittent,intermittent2}.
Another widely applicable process concerning optimal search strategies are
L\'{e}vy flights , which are based on random walk processes with long- tailed
jump length distributions and are known to be an efficient strategy for
finding a target of unknown place \cite{lomholt,Viswanathan2}.  A species
which is known to move in L\'{e}vy patterns are wandering albatrosses
\cite{Viswanathan1,humphries} or marine predators as sharks, bony fishes,
sea turtles and penguins \cite{sims1,sims2}. It would thus be interesting to
study effects of self volume in these models as well.

\ack
We thank Andrey Cherstvy for discussions. We acknowledge funding through
an Alexander von Humboldt Fellowship and ARRS Program No. P1-0002 (A.G.)
and an Academy of Finland FiDiPro grant (R.M.).

\appendix

\section{Initial distance between Predator and nearest prey}

We determine the initial distance between the predator and the nearest prey
on a two dimensional square lattice with edge length $L$.  The predator
sits in the centre of the lattice and the prey are randomly distributed on
the remaining $N_S=L^2-1$ sites.  As the prey have a self volume, a lattice
site can only be occupied by a single prey.  The probability for the minimal
distance between predator and nearest prey $d_n$ to be equal to $d$ is
\begin{equation}
P(d_n=d)=P(d_n\geq d)-P(d_n\geq d+1).
\end{equation}
We calculate the probability function $p(d_n\geq d)$ using combinatorics.
If $d_n \geq d$ all sites within distance $d$ (up to distance $d-1$) must
be unoccupied.  To obtain the number of these sites we count all sites at
exactly distance $d$ and add them from distance $1$ up to $d-1$.  The number
of sites at distance $d$ can be shown to be
\begin{equation}
N(d)=\left\{\begin{array}{ll}4d, & d\leq(L-1)/2 \\
4(L-d), &  d>(L-1)/2 \end{array}\right..
\end{equation}

Counting all empty sites within the distance $d$ from the predator leads to
\begin{equation}
M(d)=\sum\limits_{i=1}^{d-1} N(i).
\end{equation}
That is explicitly,
\begin{equation}
\hspace*{-1.2cm}
M(d)=\left\{\begin{array}{ll}\sum\limits_{i=1}^{d-1} 4i=2d(d-1), & d
\leq(L+1)/2\\
N_S-\sum\limits_{i=1}^{L-d}4i=\left(L^2-1\right)-2(L-d)(L-d+1), & d>
(L+1)/2\end{array}\right.
\end{equation}
Due to the predator sitting in the centre there are in general $N_S=L^2-1$
possible sites for the prey to be placed on.  Under the assumption that
the minimal distance is $d$, i.e., $M(d)$ sites are vacant, there are
$N_R(d)=N_S-M(d)$ remaining sites for the prey.  The probability for the
minimal distance to be greater or equal $d$ is the number of possibilities to
place the prey at the remaining sites $N_R(d)$ over the possibilities to place
the prey at sites greater equal every possible distance ($1$ to $d_{\mathrm{max}}$)
\begin{equation}
P(d_n\geq d)={N_R(d_i) \choose N_P}\Big/\sum\limits_{i=1}^{d_{\mathrm{max}}}
{N_R(d_i)i \choose N_P}.
\end{equation}
For the probability function of $d_n$ using Eq.~(\ref{eq_Appendix1}) we obtain
\begin{equation}
P(d_n)=\left({N_R(d) \choose N_P}-{N_R(d+1) \choose N_P}\right)\Big/\sum\limits_{
i=1}^{d_{\mathrm{max}}} {N_R(d_i) \choose N_P},
\end{equation}
where we define $d_{\mathrm{max}}$ as the maximal possible distance between the
predator and the nearest prey.  It is determined by the number of prey
(due to the self volume of the prey) and can be calculated by allocating
all prey as greatest distance as possible starting at $d=L-1$.  Then the
first fully unoccupied diamond at distance $d$ is the maximal possible
distance $d_{\mathrm{max}}$.  There exist the following condition to place all prey
$N_P\leq N_S-M(d_{\mathrm{max}})$,
\begin{equation}
\left\{\begin{array}{ll}
N_P\leq\sum\limits_{i=1}^{L-1-d_{\mathrm{max}}}4i, & N_P<\frac{N_S}{2}\\
N_P-\frac{L^2-1}{2}\leq\sum\limits_{i=d_{\mathrm{max}}}^{\frac{L-1}{2}}4i,
& N_P\geq\frac{N_S}{2}\end{array}\right..
\end{equation}
We then get the maximal possible distance as a function of prey,
\begin{equation}
d_{\mathrm{max}}=\left\{\begin{array}{ll}\left\lfloor \frac{2L-1-\sqrt{1+2N_P}}{2}
\right\rfloor, & N_P<N_S/2 \\[0.6cm]
\left\lfloor\frac{\sqrt{1+2(N_S-N_P)}-1}{2}\right\rfloor,  & N_P\geq
N_S/2 \end{array}\right.
\end{equation}
where $\lfloor x\rfloor:=\max \{ m\in\mathbb{Z} \mid m\leq x\} $ is the floor
function. We now obtain the expectation value of the initial distance from the
predator to the nearest prey,
\begin{equation}
\left\langle d_n\right\rangle=\sum\limits_{d_i=1}^{d_{\mathrm{max}}}p(d_{
\mathrm{min},i})d_{\mathrm{min},i}
\end{equation}
such that
\begin{equation}
\left\langle d_n\right\rangle=\frac{\sum\limits_{d_i=1}^{d_{\mathrm{max}}}d_i
{N_R(d_i) \choose N_P}-{N_R(d_i+1) \choose N_P}}{\sum\limits_{d_i=1}^{d_{
\mathrm{max}}}{N_R(d_i) \choose N_P}-{N_R(d_i+1) \choose N_P}}.
\end{equation}

 \begin{figure}
 \begin{center}
\includegraphics[angle=-90,width=8cm]{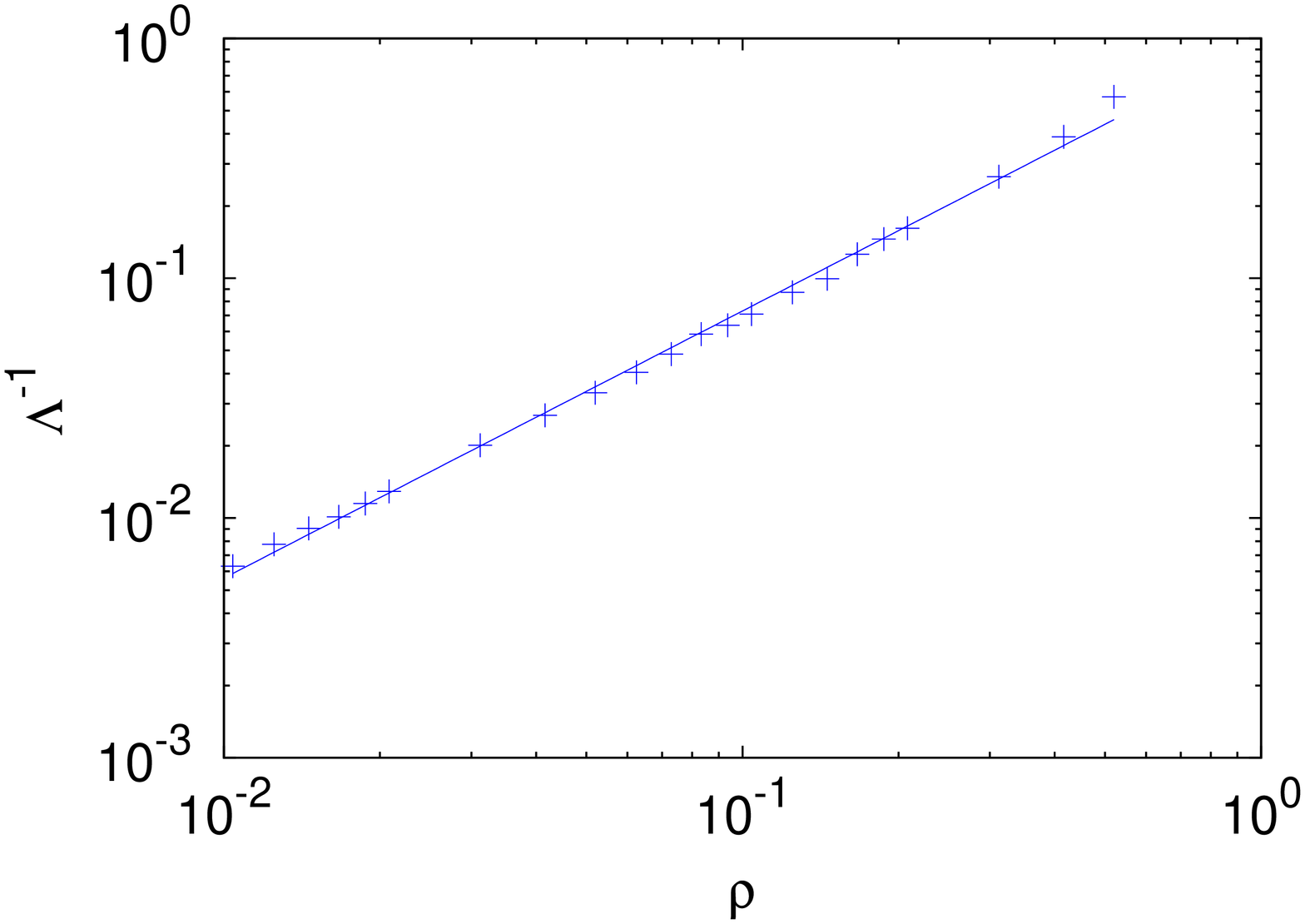}
\end{center}
\caption{Parameter $\Lambda$ from the exponential fits in Fig.~\ref{FP} as
function of prey density.}
\label{fig_lambda}
\end{figure}

\section{Exponents of Fig.~\ref{fig:CaptureTime}, Fig.~\ref{FP} and
Fig.~\ref{SpringConstants}}
\label{appb}

We here present plots depicting the dependence of the parameter $\Lambda$ from
Fig.~\ref{FP} versus the prey density (Fig.~\ref{fig_lambda}) as well as of the
scaling exponents $\beta$ and $\nu$ from Figs.~\ref{fig:CaptureTime} and
\ref{SpringConstants}.

\begin{figure}
\includegraphics[angle=-90, width=8cm]{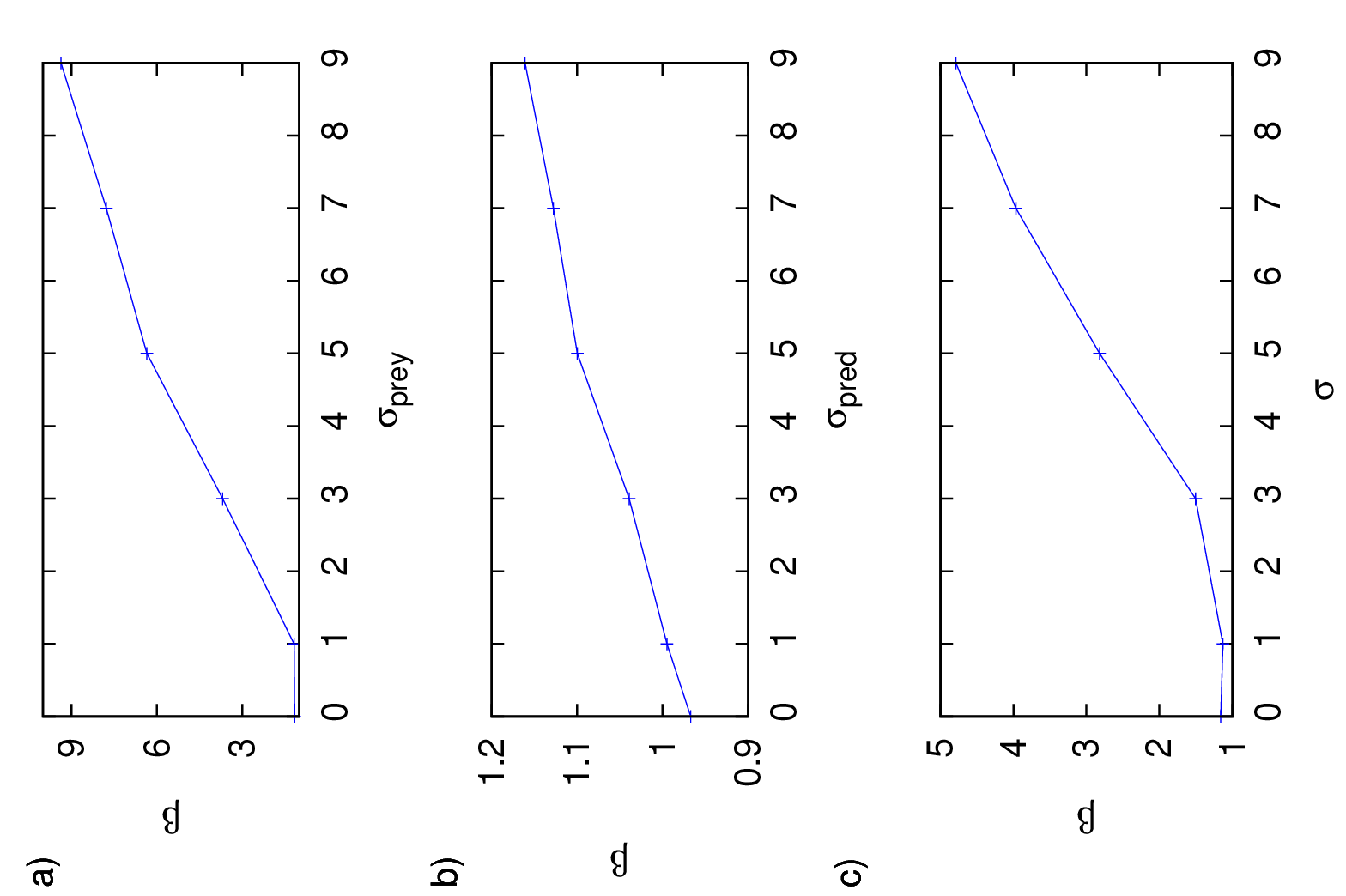}
\hfill
\includegraphics[angle=-90, width=8cm]{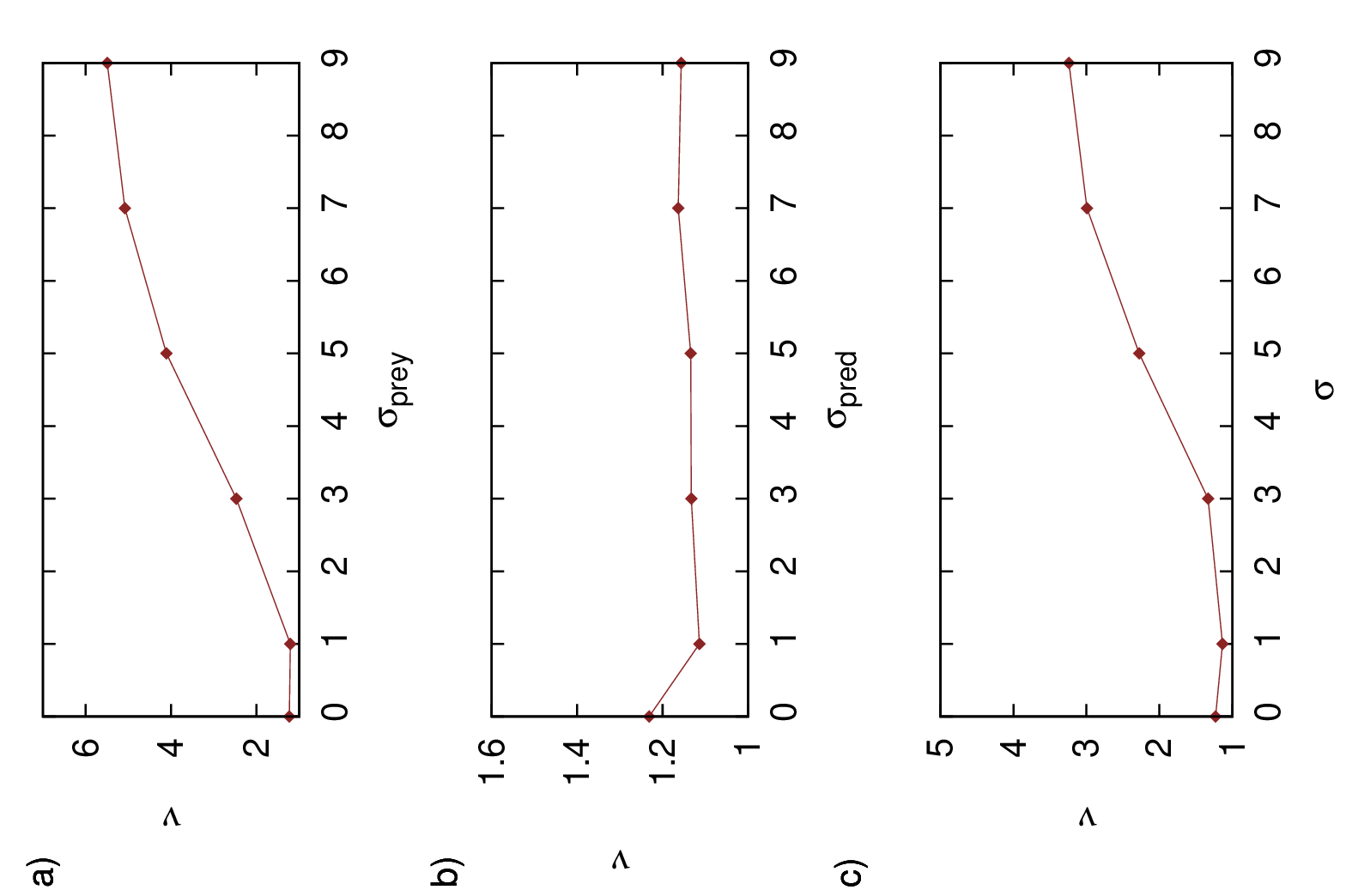}
\caption{Left: Exponents of the power-law fits in Fig.~\ref{fig:CaptureTime}
a) as function of the preys' sighting range in case of a blind predator, 
b) as function of the predator's sighting range in case of blind prey, 
c) as function of the sighting range in case of identical sighting ranges
.
Right: Exponents of the power-law fits in Fig.~\ref{SpringConstants}
a) as function of the preys' sighting range in case of a blind predator, 
b) as function of the predator's sighting range in case of blind prey, 
c) as function of the sighting range in case of identical sighting ranges
. The lines are meant to guide the eye.}
\label{fig_beta_nu}
\end{figure}

\end{document}